# Crystallographic and magnetic structures of the VI$_3$ and LiVI$_3$ van der Waals compounds

Thomas Marchandier[1,2,3], Nicolas Dubouis[1,2,3], François Fauth[4], Maxim Avdeev[5,6], Alexis Grimaud[1,2,3], Jean-Marie Tarascon[1,2,3] and Gwenaëlle Rousse[1,2,3]

1. Collège de France, Chaire de Chimie du Solide et de l'Energie, UMR 8260, 11 place Marcelin Berthelot, 75231 Paris Cedex 05, France

2. Réseau sur le Stockage Electrochimique de l'Energie (RS2E), FR CNRS 3459, 75005 Paris, France

3. Sorbonne Université– 4 place Jussieu, F-75005 Paris, France

4. CELLS -ALBA synchrotron, Cerdanyola del Valles, Barcelona E-08290, Spain

5. School of Chemistry, the University of Sydney, Sydney, NSW 2006, Australia

6. Australian Centre for Neutron Scattering, Australian Nuclear Science and Technology Organisation, New Illawarra Rd, Lucas Heights, NSW 2234, Australia

**Abstract :**

Two-dimensional (2D) layered magnetic materials are generating a great amount of interest for the next generation of electronic devices thanks to their remarkable properties associated to spin dynamics. The recently discovered layered $VI_3$ ferromagnetic phase belongs to this family, although a full understanding of its properties is limited by an ill-defined crystallographic structure. This is not any longer true. Here, we investigate the $VI_3$ crystal structure upon cooling using both synchrotron X-ray and neutron powder diffraction and provide structural models for the two structural transitions occurring at 76 K and 32 K. Moreover, we confirm by magnetic measurements that $VI_3$ becomes ferromagnetic at 50 K and discuss the difficulty of solving its full magnetic structure by neutrons. We equally determined the magnetic properties of our recently reported $LiVI_3$ phase, which is alike the well-known $CrI_3$ ferromagnetic phase in terms of electronic and crystallographic structures and found to our surprise an antiferromagnetic behavior with a Néel temperature of 12 K. Such a finding provides extra clues for a better understanding of magnetism in these low dimension compounds. Finally, the easiness of preparing novel Li-based 2D magnetic materials by chemical/electrochemical means opens wide the opportunity to design materials with exotic properties.

**Introduction**

2D materials such as graphene are among the most promising candidates to enhance electronic devices while enabling future upcoming technologies, hence the great attention that they are capturing in the scientific community [1]. Moreover fundamental-wise, they offer a fascinating playground for physicists owing to the complexity of their spin dynamics involving ferromagnetic and antiferromagnetic resonances[2]. For instance, although in an ideal magnetically isotropic 2D material long range magnetic order is impossible due to the Mermin-Wagner theorem [2], the discovery of ferromagnetic order in two-dimensional materials such as $Cr_2Ge_2Te_6$ [3] or $CrI_3$ [4] opened new horizons to investigate spintronic systems. Thus, the discovery of new ferromagnetic-layered van der Waals (vdW) materials is of prime importance as illustrated by the recent discovery of ferromagnetic order in the layered semi-conductor $VI_3$ [5].

$VI_3$ presents a Curie temperature slightly lower than the widely studied $CrI_3$ (50 K vs 60 K) but a different electronic structure ($d^2$ vs $d^3$). The two compounds $VI_3$ and $CrI_3$ provide a unique opportunity to understand the implication of the electronic structure on their magnetic properties. However, this could not happen yet because of an ill-defined crystal structure that is still subject of intense debates and the source of contradictory reports [5–7] for the room temperature structure. Moreover, upon cooling, $VI_3$ undergoes two-successive phase transitions at around 76 K [5,7] and 32 K [8] that are so far orphaned of structural model. Additionally, experimental and theoretical investigations suggest quite complex magnetic behavior for $VI_3$, with namely a ferromagnetic transition that occurs nearly midway (~ 50 K) between the structural transitions whose magnetic structure has not been yet established.

In this paper, we investigate both the crystal and magnetic structure of $VI_3$ from room temperature (RT) down to 3 K using high-angular resolution synchrotron X-ray powder diffraction (SXRPD) and neutron powder diffraction. In particular, we confirm the $R\bar{3}$ structural model at room temperature and the presence of two structural transitions at low temperature around 76 K and 32 K. For the first time, we solve the crystal structures of low temperature phases of $VI_3$ and discuss its magnetic structure with the help of both low temperature neutron diffraction and magnetization measurements. In parallel, we measured the magnetic behavior of the new layered $LiVI_3$ phase, recently synthesized by our group via chemical reduction [9] and found an antiferromagnetic behavior as compared to the ferromagnetic behavior reported for $CrI_3$ [3] , although both compounds are alike in terms of structural and electronic configurations.

## II Results and discussion

### A. Experimental procedures:

Pure VI$_3$ powders were obtained through iodine vapor transport in quartz tubes under high vacuum as already described elsewhere [5]. Vanadium metal powder (Alfa Aesar, 99.9 %) was mixed with a slight excess of iodine (1:1.1 molar ratio) inside an Argon filled glovebox, the mixture was then poured in a quartz tube subsequently sealed under high vacuum (10$^{-6}$ bar). The tube was placed in a tubular furnace in such a way that one end of the tube was at room temperature whereas the other one was positioned in the middle of the furnace and heated at 450 °C for 72 hrs. After reaction, the tube was opened inside a glovebox and the powder was placed in a Schlenk flask and purified under dynamic vacuum at 200 °C. LiVI$_3$ powder was prepared by stirring VI$_3$ powder in an excess (1:3 molar ratio) of n-Butyl Lithium (2.5 M in hexane, Sigma Aldrich) under Argon atmosphere. After 1-hour reaction, the suspension was centrifuged and rinsed three times with hexane before being dried under dynamic vacuum.

Synchrotron X-ray powder diffraction (SXRPD) patterns were collected on the BL04-MSPD beamline of the ALBA synchrotron (Barcelona area) using the Multi Analyzer Detection (MAD) setup offering the highest possible instrumental angular resolution, Δ(2Θ) < 0.006° in the 0-35° 2Θ range [10,11]. Powders were sealed in Ar glove box in a 0.6 mm diameter borosilicate capillary and transferred to the Dynaflow liquid helium flow cryostat. [12] SXRPD experiments were performed at 32 keV energy, slightly below the I K edge to minimize absorption. The exact wavelength, (λ = 0.38692 Å), was determined from the six first Bragg reflections of a NIST silicon standard SRM 640e.

Constant wavelength neutron powder diffraction (NPD) data were collected on the ECHIDNA high-resolution and WOMBAT high-intensity neutron powder diffractometers [13,14] using the wavelengths of 2.4395 Å and 2.4184 Å, respectively. For VI$_3$, the NPD data were collected on ECHIDNA at 9, 40, 60, and 90 K and on WOMBAT every 1 K in the range 3-92 K. For LiVI$_3$, the NPD data were collected only on ECHIDNA at 3 and 25 K. To prevent reaction of the samples with ambient atmosphere, the samples were loaded into 9 mm diameter cylindrical vanadium cans in Ar glove box and sealed with an In wire.

Finally, all the SXRPD diffraction patterns were refined using the Rietveld method with the FullProf program [15].

The evolution of the magnetic susceptibility with temperature was measured using a SQUID (XL, Quantum Design), under zero field cooled (ZFC) conditions between 2 K and 400 K with an applied magnetic field of 10 kOe. Magnetization curves were recorded at 2 K varying magnetic field in a [-70 kOe; 70 kOe] range.

B. Crystal structure of VI$_3$ from RT to 9 K

*a. Room temperature structure*

VI$_3$ have long been reported to adopt a BI$_3$-type structure [16] analogous to VCl$_3$ and VBr$_3$ with the $R\bar{3}$ space group, but recent publications have challenged this assignment with reports of alternatives $P\bar{3}1c$ [7] or *C*2/*m* [6] space groups. These different models commonly agree that VI$_6$ octahedra share edges to form layers into which vanadium atoms fill 2/3 of the octahedral positions to form a honeycomb pattern. These honeycomb layers are then stacked one above the other separated by a van der Walls gap. Their discrepancies are nested in the different stacking sequences of the VI$_3$ layers along the c-axis. For instance, the *C*2/*m* model proposed by Tian et al. implies an ABC stacking of the iodine atoms in opposition to an ABAB one for both the $R\bar{3}$ and the $P\bar{3}1c$ models. The difference between the two latter is more subtle and originates from different honeycomb stackings, which result in only minute variations in the X-ray diffraction patterns. In order to clarify the VI$_3$ room temperature structure, we tried to refine the experimental synchrotron powder pattern collected at room temperature with those different models (cf. Figure S1). While the *C*2/*m* model is clearly inadequate to index our pattern, the other models led to decent fits, and solely the indexation of small low angle reflections specific to the vanadium superstructure allowed discriminating them. The only space group able to index all these peaks was $R\bar{3}$, with lattice parameters *a* = 6.9277(2) Å, *c* = 19.9389(2) Å and this model led to the best refinement (Figure 1a and b and Table S1). In this structure, the iodine layers are stacked in an ABAB manner (O1 type) whereas the vanadium honeycomb motifs adopt an ABC sequence.

It is worth mentioning that a good refinement can also be obtained in a nine-fold smaller unit cell with *a* = 3.9998(2) Å, *c* = 6.6462(2) Å ($P\bar{3}m1$ space group) (cf. Figure S1 and Table S2). This cell indexes all peaks, except the superstructure ones associated to the honeycomb ordering, and corresponds to a structural model in which vanadium sits in a 1*a* Wyckoff site with a 2/3 occupancy. For the rest of the manuscript this model will be referred as "the average model" as opposed to the "complete model" taking into account the honeycomb vandium superstructure.

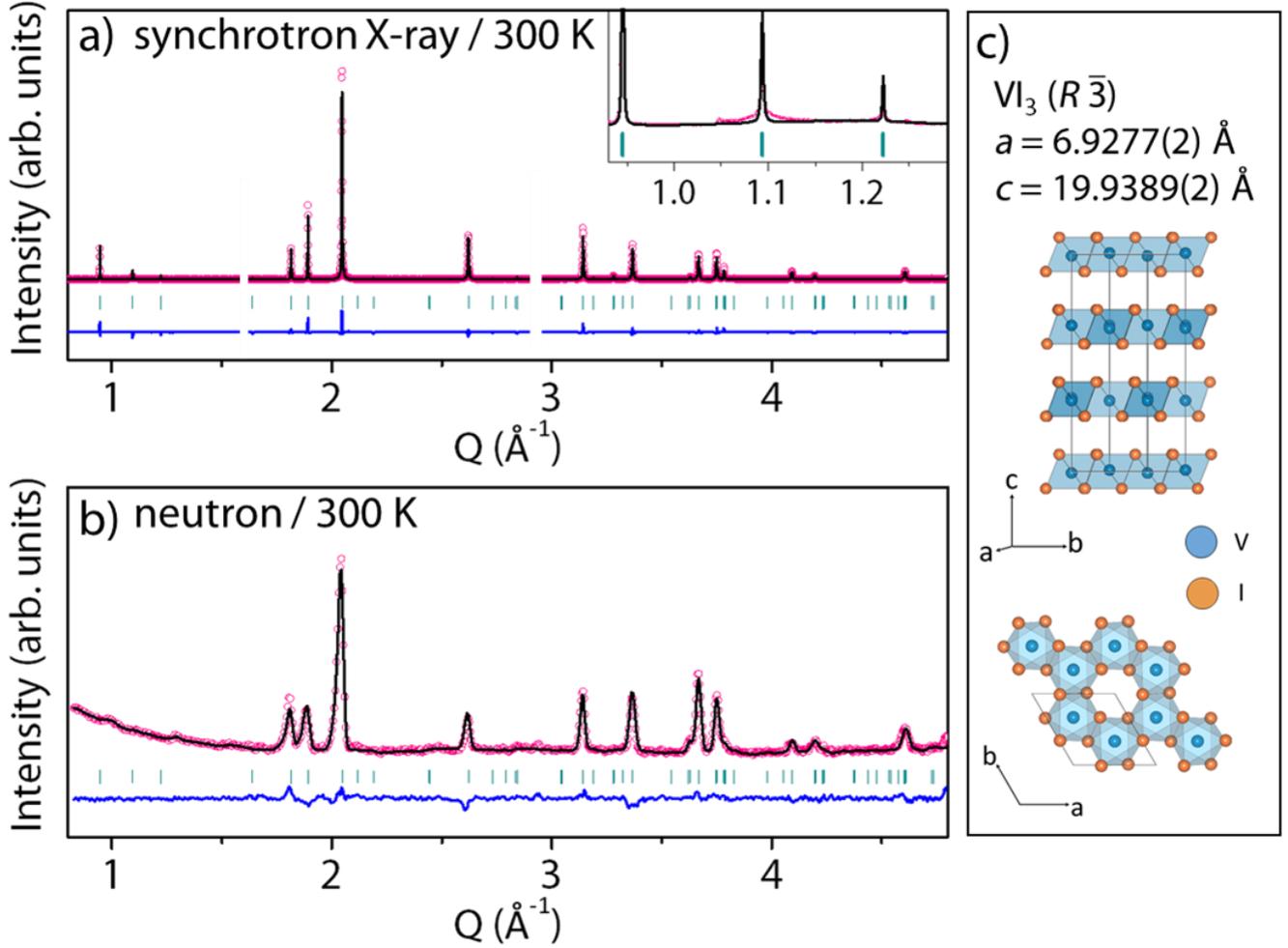

**Figure 1**: Synchrotron X-ray (a) and neutron (b) Rietveld refinements of $VI_3$ at 300 K. Wavelengths for synchrotron X-ray and neutrons are $\lambda = 0.3869$ Å and $\lambda = 2.4184$ Å, respectively. The pink circles, black continuous line, and bottom blue line represent the observed, calculated, and difference patterns, respectively. Vertical green tick bars stand for the Bragg positions. Two regions were excluded due to minute amount of impurities. The inset in Figure a) highlights the indexation of the superstructure peaks c) Structure of $VI_3$ at 300 K.

b. *Low temperature structures*

The structural evolution of $VI_3$ upon cooling was explored by high-angular resolution synchrotron X-ray powder diffraction. From RT to 80 K the structure described above is preserved with however a contraction of the lattice parameters. When lowering the temperature further, the SXRD patterns shows a first abrupt splitting of peaks (cf. Figure 2) at 76 K and a second one less drastic at temperatures around 32 K. Such features are in agreement with the reported first and second order phase transitions at $T_1 \approx 76$ K [5–8] and $T_2 = 32$ K [8] respectively. Nevertheless and to the best of our knowledge, no structural model has ever been proposed for the lowest temperature phase. For the intermediate phase, the only model reported in literature [7] was deduced from a X-ray pattern showing a strong preferred orientation, and whose RT analog was refined in $P\bar{3}1c$, a space group we excluded as discussed earlier. Thus, we embarked on solving the structure of these two low temperature phases using high-angular resolution SXRPD patterns measured at 60 K and 9 K with a long counting time. At a first approach, the structures were solved without considering the vanadium honeycomb superstructure ("average model"), and in a second time a more complex but realistic structural description was built by adding the honeycomb vanadium superstructure to the average model ("complete model").

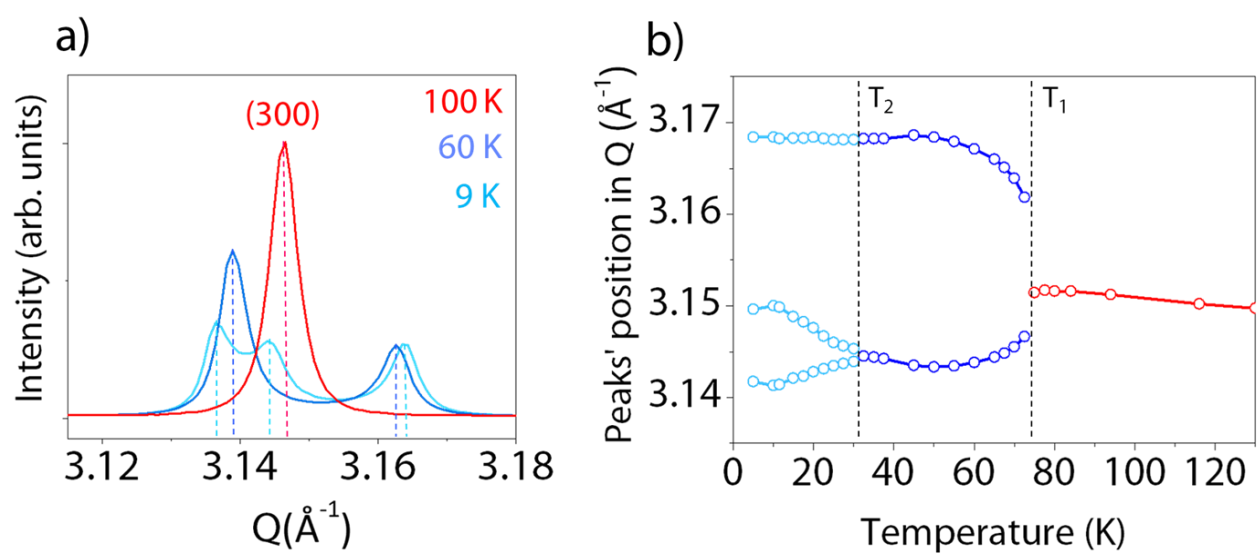

**Figure 2**: Evolution of selected reflections with temperature a) evolution of the (300) peak (in $R\bar{3}$ setting) b) evolution of the peaks' positions.

*Average model*

The 60 K SXRD pattern (excluding the superstructure peaks) can be indexed in a monoclinic unit cell with *a* = 6.9395(2) Å, *b* = 3.9671(2) Å, *c* = 6.5960(2) Å, β = 90.4539(4)°, that is related to the RT average cell using the following transformation: $(\mathbf{a_m}\ \mathbf{b_m}\ \mathbf{c_m}) = (\mathbf{a_h}\ \mathbf{b_h}\ \mathbf{c_h})\begin{pmatrix} 1 & 1 & 0 \\ 0 & 2 & 0 \\ 0 & 0 & 1 \end{pmatrix}$

Where the $m$ and $h$ subscripts stand for monoclinic and hexagonal ($P\bar{3}m1$ space group) cells, respectively. Then, the pattern was refined via the Rietveld method using the space group *C*2/*m* confirming the plausibility of the model (cf. Figure S2 and Table S3). At 9 K, the symmetry was further lowered to the $P\bar{1}$ space group without changing the unit cell metrics and the model was refined against the SXRPD pattern (cf. Figure S2 and Table S4). A decent fit of the SXRPD pattern, which gave a satisfactory representation of the structure was obtained with the average model, and questions the necessity to further refine the model. However, although contributing only to a limited extent to the diffraction pattern intensity, the superstructure peaks are of prime importance to understand the physical properties of the material and have to be considered.

|  | 9 K | 60 K | 300 K |
|---|---|---|---|
| Average model (without superstructure) | $P\bar{1}$<br>a = 6.9353(2) Å  α = 90.1044(6)°<br>b = 3.9655(2) Å  β = 90.3956(5)°<br>c = 6.5910(2) Å  γ = 90.1495(5)°<br>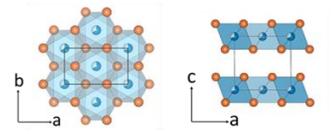 | $C2/m$<br>a = 6.9395(2) Å  β = 90.4539(4)°<br>b = 3.9671(2) Å<br>c = 6.5960(2) Å<br>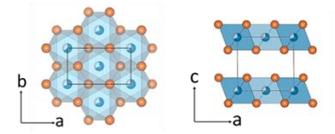 | $P\bar{3}m1$ (Hexagonal cell)<br>a = 3.9998(2) Å<br>c = 6.6462(2) Å<br>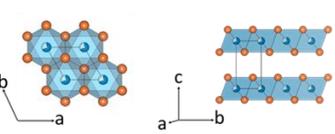 |
| Complete model (with superstructure) | $P\bar{1}$<br>a = 7.7268(2) Å  α = 53.1280(2)°<br>b = 7.6808(2) Å  β = 53.0863(3)°<br>c = 7.6985(2) Å  γ = 53.5022(3)°<br>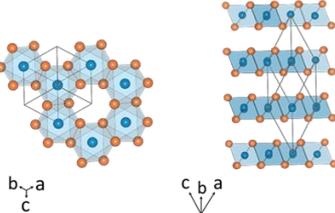 | $P\bar{1}$<br>a = 7.7359(2) Å  α = 53.1933(5)°<br>b = 7.6886(2) Å  β = 53.0168(7)°<br>c = 7.6968(2) Å  γ = 53.4732(7)°<br>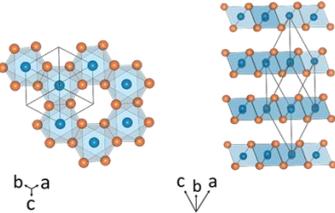 | $R\bar{3}$ (Hexagonal cell)<br>a = 6.9277(2) Å  c = 19.9389(2)°<br>$R\bar{3}$ (Rhombohedral cell)<br>a = 7.7570(2) Å  α = 53.0447(2)°<br>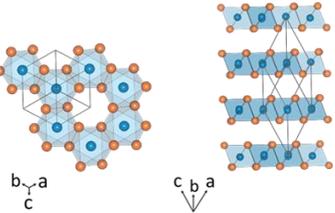 |

**Figure 3**: Overview of the different structural models at 9 K, 60 K and 300 K. The upper raw presents the "average models" where the honeycomb vanadium superstructure is not taken into account. The lower raw presents the "complete models" including the vanadium superstructure.

*Complete model*

Assuming that the vanadium superstructure observed at RT (honeycomb layers stacked in a ABC sequence) is preserved in the low temperature phases, a supercell was built from the average model (cf. SI section 2b for details) with the following transformation:

$$(a_s \quad b_s \quad c_s) = (a_a \quad b_a \quad c_a) \begin{pmatrix} -1/2 & 1/2 & 0 \\ 1/2 & 1/2 & -1 \\ 1 & 1 & 1 \end{pmatrix}$$

Here the *s* and *a* subscripts stand for supercell and average cell, respectively. At room temperature, this supercell represented in Figure 3 corresponds to the rhombohedral cell of the $R\bar{3}$ model (a = b = c = 7.7570(2) Å and α = β = γ = 53.0447(2)°). Then, the Rietveld refinements of the 60 K and 9 K patterns using this model led to a good indexation of the superstructure peaks splitting observed on cooling (cf.

Figure 4 and Table S5 and S6). The visible shape mismatch between the calculated and the measured superstructure peaks likely accounts for the presence of stacking faults as commonly encountered in such layered materials [17] and strain developing across the phase transitions. Finally, this two-step refinement process enabled to derivate the atomic positions and lattice parameters from the average model which prevented our refinements to diverge with such a large amount of refined parameters.

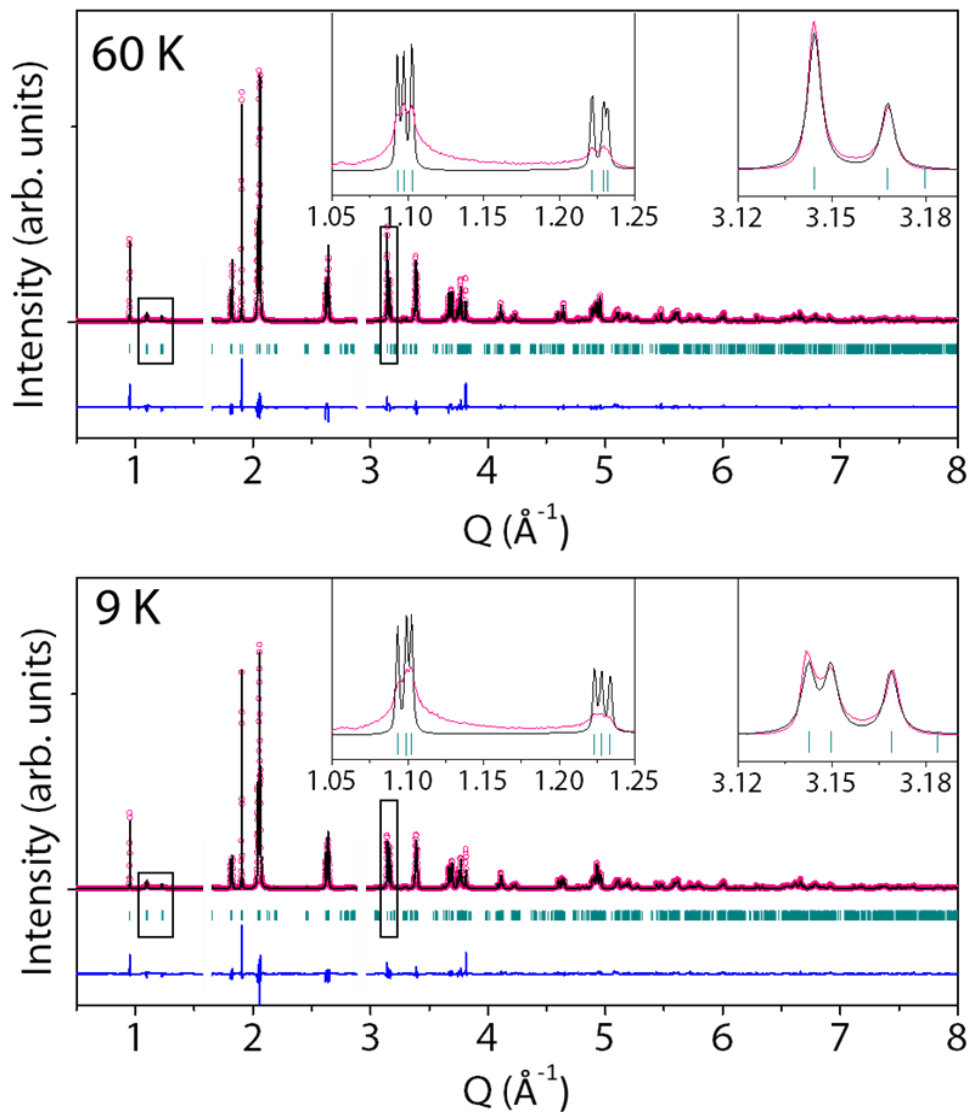

**Figure 4**: Synchrotron XRD Rietveld refinement of VI$_3$ at 60 K (top) and 9 K (bottom) using the "complete model" in $P\bar{1}$ space group reported in Tables S5 and S6. Wavelength for synchrotron X-ray is λ = 0.3869 Å. The pink circles, black continuous line, and bottom blue line represent the observed, calculated, and difference patterns, respectively. Vertical green tick bars stand for the Bragg positions. Some regions were excluded due to minute amount of impurities. The insets in each figure refer to a zoom of the angular regions shown by a black rectangle in the main Figure. The middle inset highlights the indexation of the superstructure peaks whereas the right one highlights the Q range represented in Figure 2a.

To further understand the origin of the observed distortions we represent in Figure 5 the evolution of the V-V distances in the honeycomb layers. Above 76 K the vanadium atoms describe regular hexagons in the layers. In contrast, below this temperature, a contraction of the of the V-V distances along a single direction leads to the formation of vanadium dimers. This finding contradicts the vanadium anti-dimerization deduced by Son et al [7]. Finally, below 32 K, the honeycomb framework is distorted in an additional direction leading to three distinct V-V distances inside the layer. Similar behaviors have been reported for other transition metal trichlorides $MCl_3$ (namely $TiCl_3$, $MoCl_3$ or $TcCl_3$), with however a larger difference between long and short M-Cl distances [18].

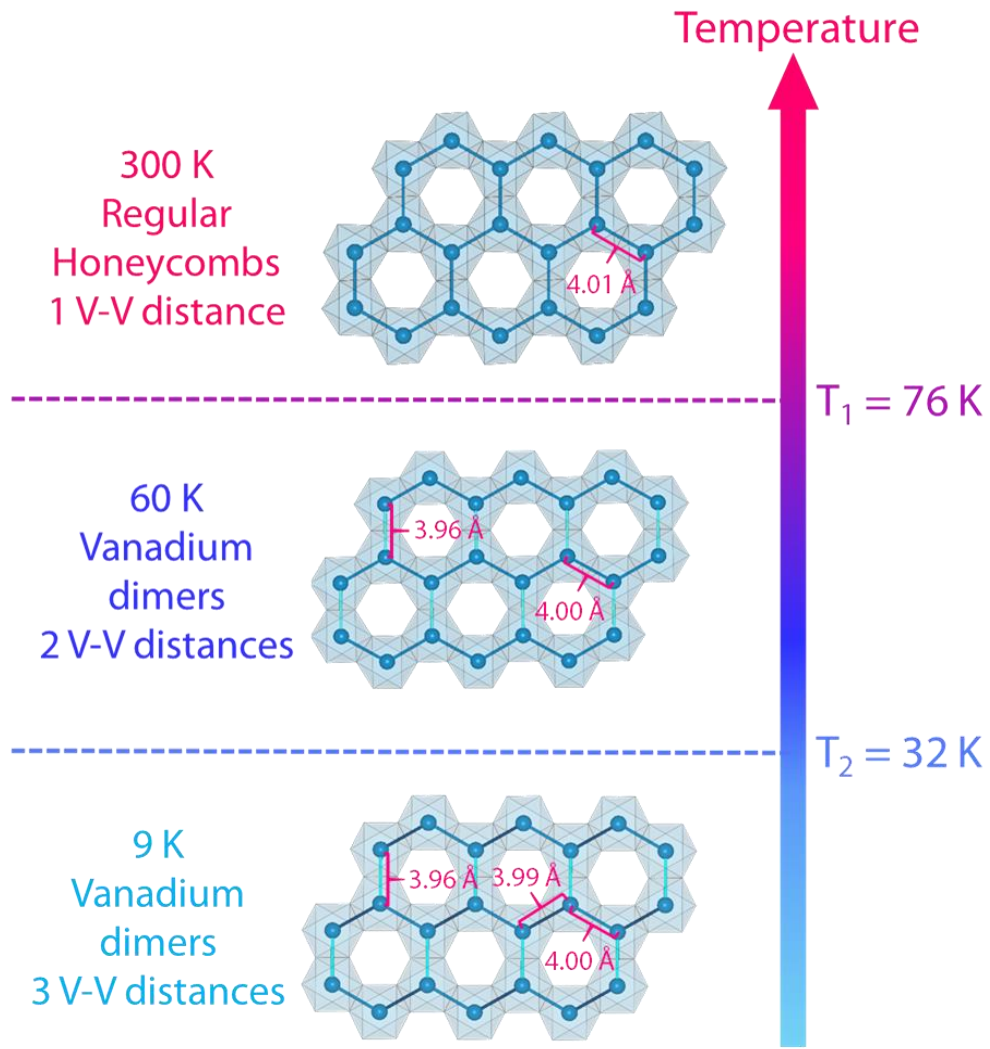

**Figure 5**: Evolution of the V-V distances in the honeycomb motifs at 300 K, 60 K and 9 K, deduced from the complete structural models reported in Table S1, S5 and S6.

At this stage, it is important to consider the connection between those complex structural transitions and the onset of ferromagnetic order at $T_C$ = 50 K in $VI_3$ that we confirm in our homemade sample (cf. following section). Obviously, the 76 K structural phase transition that occurs above $T_C$ (50 K) is free of magnetic effects. Between $T_1$ and $T_2$ it is worth noting that the evolution of the lattice parameters and characteristic distances in the $VI_3$ structure are not monotonous (cf. Figure 6). More specifically, we observe that the V-V and interlayer (van der Walls) distances (cf. Figure 6 a and b) increase from 76 K to 50 K while they decrease upon further cooling. This anomaly is directly linked to the establishment of a magnetic order in the structure and is strongly indicative of a magnetostriction phenomenon already suggested by Dolezal et al [8]. Finally, the origin of the low temperature structural phase transition at $T_2$ = 32 K still remains to be explained in conjunction with the magnetic ordering at 50 K.

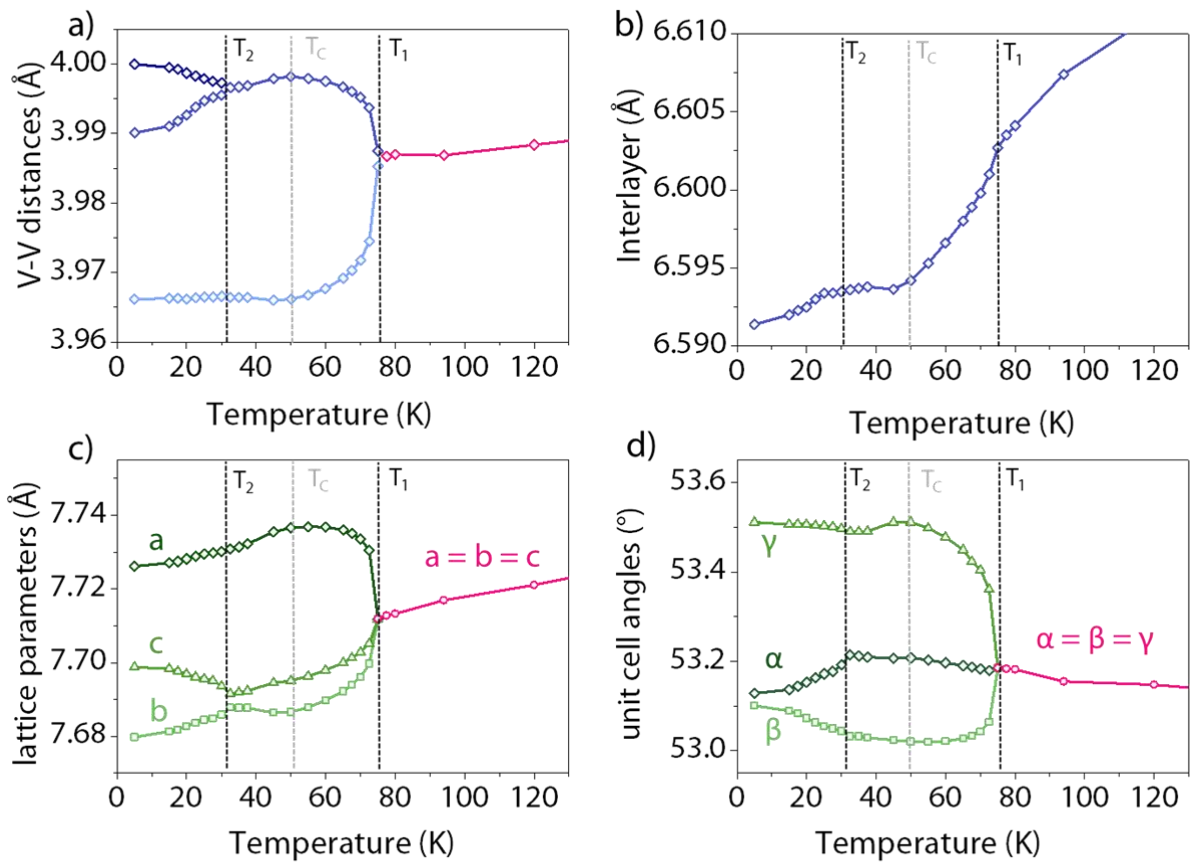

**Figure 6**: Evolution of a) the vanadium-vanadium distances, b) the interlayer distance (defined as the distance between two vanadium layers) and c), d) lattice parameters of VI$_3$ from 10 K to 140 K. These parameters are given using the complete model settings described in Figure 3.

To interrogate the presence of a long-range magnetic ordering in VI$_3$, we collected neutron diffraction patterns between 3 K and 100 K. The patterns are reported in Figure 7a and confirm the occurrence structural phase transitions at T$_1$ = 76 K and T$_2$ = 32 K previously spotted by X-rays. The magnetic signature related to the ferromagnetic ordering is in contrast much more difficult to observe. Nevertheless some hints can be spotted by comparing the 60 K (above T$_c$), 40 K and 9 K (both below T$_c$) patterns. They show a subtle intensity evolution of the (110) peak (at Q = 1.84 Å$^{-1}$, here (hkl) indices refer to the $R\bar{3}$ hexagonal cell) (cf. Figure 7c and 7d) with no further visible evolution of this peak's intensity down to the lowest measured temperature. This observation is consistent with a ferromagnetic order with magnetic moments oriented perpendicular to the honeycomb vanadium layers (along [001]) (cf. Figure 7b), as deduced from simulations of ferromagnetic contributions along different directions (cf. SI part 3, Figure S5). However, if of magnetic origin, this subtle intensity increase would correspond to an ordered magnetic moment around 0.2 $\mu_B$ which is ten-fold smaller than expected for V$^{3+}$ (2 $\mu_B$, S = 1 for V$^{3+}$). Such an observation indicates that the macroscopic ferromagnetic ordering along [001] evidenced by magnetic susceptibility on single crystals [5] does not translate into a long-range complete ordering. It is likely that this almost-undistinguishable ordered moment arises from a reduced coherence length in the magnetic structure, which may be due to the presence of defects in the vanadium honeycomb pattern and stacking faults and weakness of the magnetic interactions across the van der Waals gap. Another possible explanation to account for the weak magnetic moment is nested in the feasibility of unpaired electrons to get paired because of a certain structural distortion, where owing to the formation of V-V dimers part of the d electrons get paired because of metal-metal bonds.

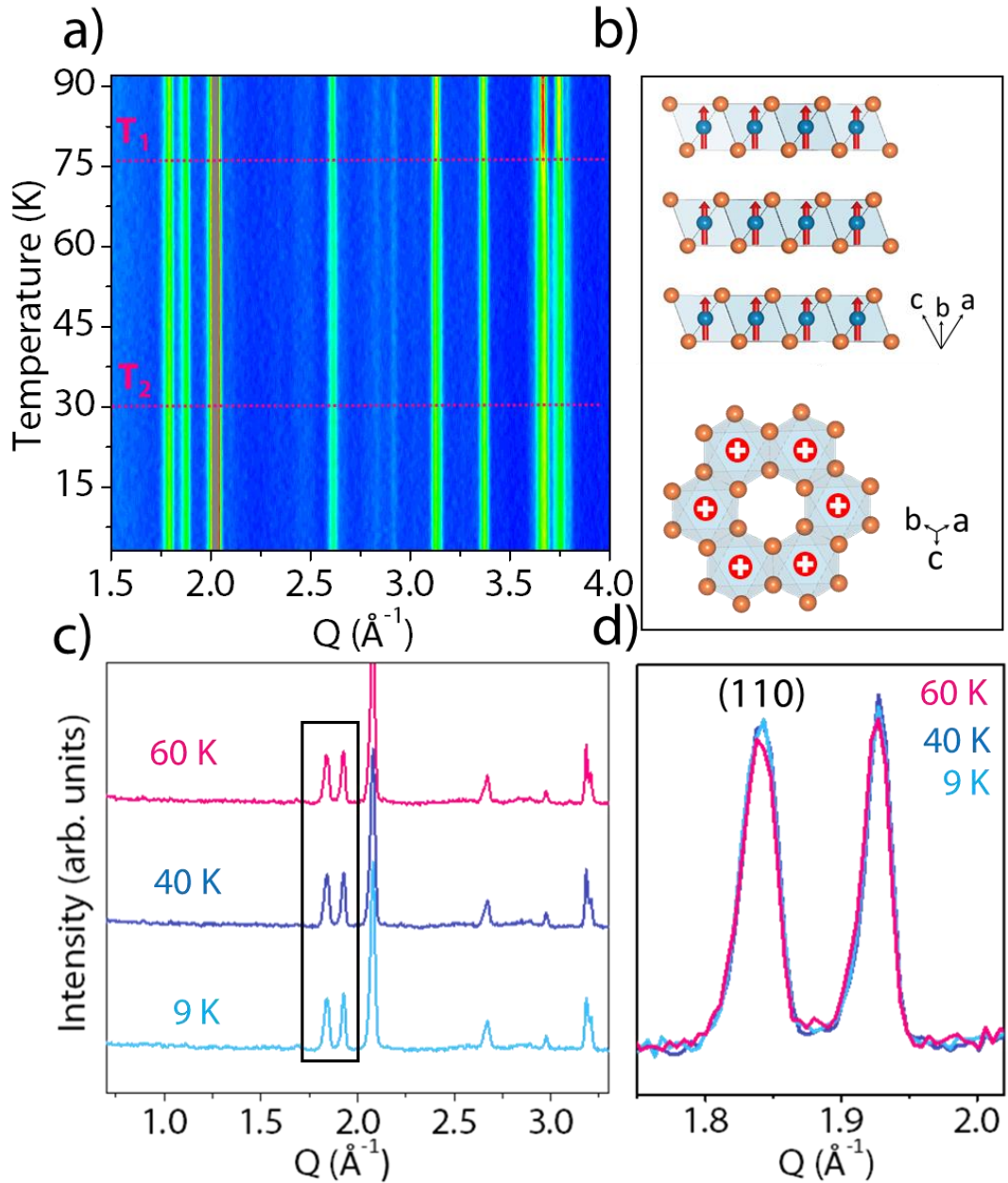

**Figure 7**: a) Evolution of the neutron diffraction patterns with temperature between 3 K and 92 K b) Magnetic structure of VI$_3$ (vanadium and Iodine atoms are in blue and orange respectively). The spin orientations are represented with red arrows or by a '+' sign indicating that the spins are pointing orthogonally to the view. c) Comparison of the 60 K, 40 K and 9 K neutron diffraction pattern in the [0.7 Å$^{-1}$; 3.3 Å$^{-1}$] Q range, d) zoom in the [1.7 Å$^{-1}$; 2.05 Å$^{-1}$] Q range to highlight the magnetic contribution to the (110) reflection (with (hkl) indices referring to the $R\bar{3}$ hexagonal cell).

C. Magnetic structure and properties of LiVI$_3$

The magnetic properties of 2D vdW magnets have been proven to strongly dependent on a plethora of parameters such as electrostatic doping [19], external pressure [20] or layer stacking sequence [21], hence an impetus to explore means of varying T$_c$ in VI$_3$. Recently, our group recently succeeded in inserting lithium atoms inside VI$_3$, by means of reducing chemically agents, to form a novel LiVI$_3$ phase. This new compound is isostructural to VI$_3$ ($R\bar{3}$, $a$ = 7.1442(2) Å, $c$ = 20.6666(2) Å) [9] but it differs in terms of electronic structure by containing one extra valence electron. Consequently, it constitutes an ideal playground to explore the electronic-magnetic relationship in 2D layered iodides.

Magnetic susceptibility measurements were performed on LiVI$_3$ from 2 K to 400 K (Figure 8d) and on VI$_3$ for stake of comparison. First, in line with precedent investigations [5–8] VI$_3$ a ferromagnetic order at around 50 K (cf. Figure 8a). The interpolation of the high temperature part of the susceptibility with a Curie-Weiss law indicates an effective moment of 2.28 µB per vanadium atom, slightly lower than the one expected for the spin configuration of V$^{3+}$ (2.83 µB) and an interpolated Weiss temperature of +43.7 K (cf. Figure 8b) suggesting predominant ferromagnetic interactions. Moreover, susceptibility measurements curves were collected at 2 K in a [-70 kOe; 70 kOe] range and demonstrate a clear hysteresis loop accounting for a 10 kOe coercitive field (cf. Figure 8c), which confirmed the ferromagnetism in VI$_3$. Surprisingly, LiVI$_3$ shows an antiferromagnetic order below T$_N$ = 12 K with a Curie Weiss temperature ϴ = -18.7 K and an effective moment of 3.55 µ$_B$ per vanadium, (ie a value a bit lower than the 3.87 µ$_B$ expected for a V$^{2+}$ atom). To shed lights on the magnetic ordering of LiVI$_3$, both synchrotron X-ray and neutron diffraction experiments were performed at low temperature. The refinement of the SXRPD pattern of LiVI$_3$ at 10 K (cf. Figure S6) indicates that the RT structure is preserved, but with solely a contraction of the lattice parameters. Lastly, the neutron diffraction patterns collected between 25 K and 3 K shows the emergence of extra magnetic peaks at low angle (cf. Figure 9b).

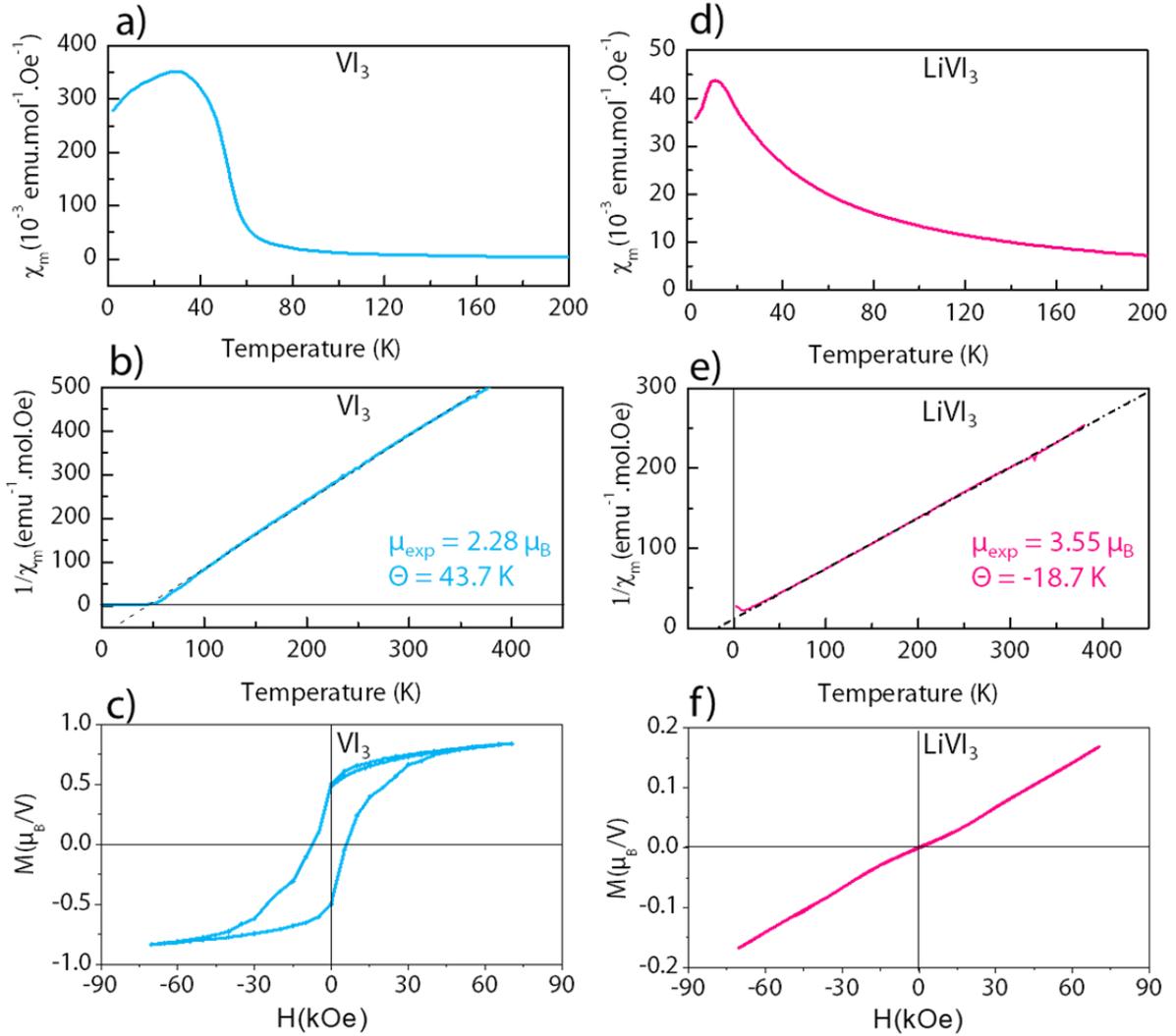

**Figure 8**: Temperature dependence of the magnetic susceptibility χ of VI$_3$ (a), and LiVI$_3$ (d). Temperature dependence of the inverse of the magnetic susceptibility 1/χ of VI$_3$ (b) and LiVI$_3$ (e), measured in zero-field-cooled (ZFC) conditions. The curves follow at high temperature an ideal Curie-Weiss behavior represented by the dashed line (CW fit). Magnetization curve of VI$_3$ (c) and LiVI$_3$ (f) as a function of the applied field measured at 2 K.

Altogether, these observations combined with the magnetic behavior described above, suggest the formation of a long-range magnetic structure at low temperature. Further exploiting the NPD patterns, we could index the extra reflections using a **k** = (0, 0, 0) propagation vector and determine the associated magnetic structure using the Isodistort program [22,23]. The 3 K neutron diffraction pattern was successfully refined in the $R\bar{3}'$ magnetic space group (cf. Figure 9d and Table S7), where all the magnetic moment are found to be oriented in the [001] direction with both intra and inter layer being antiferromagnetic coupled (cf. Figure 9c). Surprisingly, the refined magnetic moment was found to be equal to 1.68(6) $\mu_B$ i.e. lower that the expected 3 $\mu_B$ value for a $d^3$ ion. As for the VI$_3$ case, such a difference could arise from a reduced coherence length induced by defect such as stacking faults and weakness of interlayer interaction.

Understanding this Li-driven ferromagnetic to antiferromagnetic transition calls for a comparison of both VI$_3$ and LiVI$_3$ in terms of crystal and electronic structures. Firstly, the redox state of vanadium changes from V$^{3+}$ (d$^2$) to V$^{2+}$ (d$^3$) leading to half-filled t$_{2g}$ electronic states i.e. similarly to that of ferromagnetic CrI$_3$. Besides, although CrI$_3$ and LiVI$_3$ share exactly the same $R\bar{3}$ structure at low temperature, they differ by longer M-I bond lengths as well as larger interlayer distances (V-I = 2.90 Å , interlayer = 6.84 Å, at 100 K) for LiVI$_3$ as compared to CrI$_3$ (Cr-I = 2.73 Å , interlayer = 6.60 Å, at 90 K). Moreover, let's recall that several first principle calculation investigations [21,24] on bulk CrI$_3$ have revealed a strong intralayer ferromagnetic exchange as opposed to a weaker ferromagnetic interlayer coupling. Thus, presently, the increase of the interlayer and M-M distances (where M stands for transition metal) from CrI$_3$ to LiVI$_3$ may change the competing nature of the interlayer and interlayer couplings and consequently explain the antiferromagnetic interaction observed for LiVI$_3$. However, further theoretical investigation are needed to support our explanation and to better understand how the intra *vs* inter M-M distance variation impacts the magnetic properties of layered van der Waals materials.

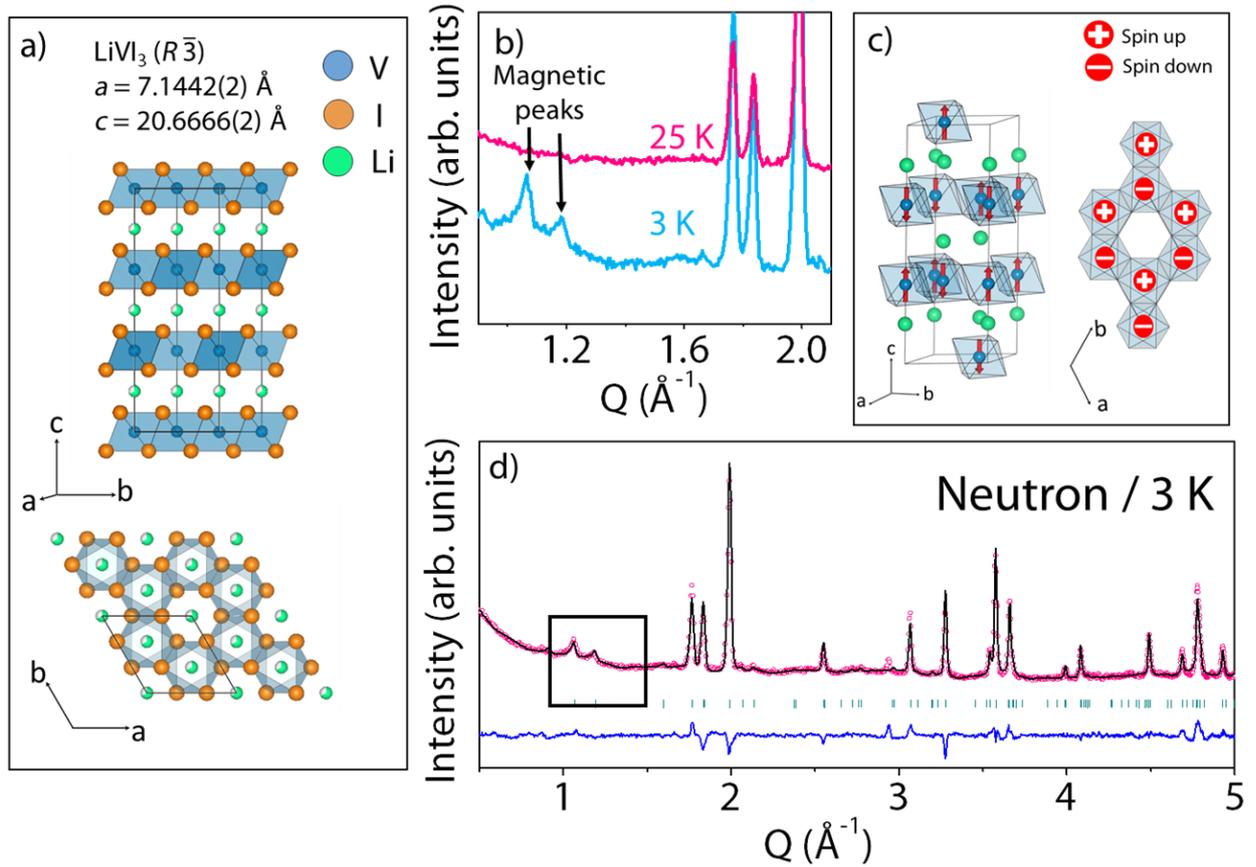

**Figure 9**: a) Structure of LiVI$_3$ at 300 K b) Neutron powder diffraction patterns of LiVI$_3$ at 25 K and 3 K c) magnetic structure of LiVI$_3$ spin orientations are represented with red arrows d) Rietveld refinements of LiVI$_3$ at 3 K (λ = 2.4395 Å). The pink circles, black continuous line, and bottom blue line represent the observed, calculated, and difference patterns, respectively. Vertical green tick bars stand for the Bragg positions.

### III Conclusion

We report here an extensive investigation of the VI$_3$ structural evolutions at low temperature using both synchrotron X-ray and neutron diffraction. We reconfirm that the room temperature $R\bar{3}$ VI$_3$ phase undergoes two structural transition upon cooling at 76 K and 32 K together with a ferromagnetic ordering at 50 K. Moreover, we provide for the first time a structural model of this material in the whole temperature range and demonstrate that the 76 K and 32 K phase transitions are solely due to distortion of the vanadium honeycomb networks with minute changes of the V-V distances. Additionally, we show by neutron diffraction that the magnetic moments are not long range ordered below T$_c$. We extended our study to the newly reported LiVI$_3$ layered phase and reveal the presence of an antiferromagnetic order at T$_N$ = 12 K in contrast to the isostructural and electronic equivalent CrI$_3$ ferromagnetic phase. We rationalize this finding that is nested in a delicate balance between the interlayer and/or the M-M distances within these materials. In this context, our work could simulate future research on finding further van der Waals ferromagnets within the halide family and beyond, hence enabling a plethora of materials for deeper comprehension of layered magnetic materials.

**Acknowledgement**

Beam time on MSPD at ALBA was granted through In-house proposal 2021024879. T.M. acknowledges the Ecole Normale Supérieure Paris-Saclay for his PhD scholarship. N.D. acknowledges the Ecole Normale Supérieure for his PhD. Scholarship. The authors acknowledge the staff of the MPBT (physical properties – low temperature) platform of Sorbonne Université for their support.

**Conflict of interest**

The authors declare no conflict of interest.

**Supplemental Material**

See Supplemental Material at **[URL will be inserted by publisher]** for additional Rietveld refinements, crystallographic tables, structure transformation details and structure files (.cif).

**Bibliography**

[1] R. Mas-Ballesté, C. Gómez-Navarro, J. Gómez-Herrero, and F. Zamora, *2D Materials: To Graphene and Beyond*, Nanoscale **3**, 20 (2011).

[2] N. D. Mermin and H. Wagner, *Absence of Ferromagnetism or Antiferromagnetism in One- or Two-Dimensional Isotropic Heisenberg Models*, Phys. Rev. Lett. **17**, 1133 (1966).

[3] C. Gong, L. Li, Z. Li, H. Ji, A. Stern, Y. Xia, T. Cao, W. Bao, C. Wang, Y. Wang, Z. Q. Qiu, R. J. Cava, S. G. Louie, J. Xia, and X. Zhang, *Discovery of Intrinsic Ferromagnetism in Two-Dimensional van Der Waals Crystals*, Nature **546**, 265 (2017).

[4] M. A. McGuire, H. Dixit, V. R. Cooper, and B. C. Sales, *Coupling of Crystal Structure and Magnetism in the Layered, Ferromagnetic Insulator $CrI_3$*, Chem. Mater. **27**, 612 (2015).

[5] T. Kong, K. Stolze, E. I. Timmons, J. Tao, D. Ni, S. Guo, Z. Yang, R. Prozorov, and R. J. Cava, *$VI_3$—a New Layered Ferromagnetic Semiconductor*, Adv. Mater. **31**, 1 (2019).

[6] S. Tian, J. F. Zhang, C. Li, T. Ying, S. Li, X. Zhang, K. Liu, and H. Lei, *Ferromagnetic van Der Waals Crystal $VI_3$*, J. Am. Chem. Soc. **141**, 5326 (2019).

[7] S. Son, M. J. Coak, N. Lee, J. Kim, T. Y. Kim, H. Hamidov, H. Cho, C. Liu, D. M. Jarvis, P. A. C. Brown, J. H. Kim, C. H. Park, D. I. Khomskii, S. S. Saxena, and J. G. Park, *Bulk Properties of the van Der Waals Hard Ferromagnet $VI_3$*, Phys. Rev. B **99**, 1 (2019).

[8] P. Doležal, M. Kratochvílová, V. Holý, P. Čermák, V. Sechovský, M. Dušek, M. Míšek, T. Chakraborty, Y. Noda, S. Son, and J. G. Park, *Crystal Structures and Phase Transitions of the Van-Der-Waals Ferromagnet $VI_3$*, Phys. Rev. Mater. **3**, 121401 (2019).

[9] N. Dubouis, T. Marchandier, G. Rousse, F. Marchini, F. Fauth, M. Avdeev, A. Iadecola, B. Porcheron, M. Deschamps, J.-M. Tarascon, and A. Grimaud, *Superconcentrated Electrolytes Insertion Electrochemistry to Layered Halides Widens Soluble*, ChemRxiv 1 (2021).

[10] F. Fauth, I. Peral, C. Popescu, and M. Knapp, *The New Material Science Powder Diffraction Beamline at ALBA Synchrotron*, Powder Diffr. **28**, (2013).

[11] F. Fauth, R. Boer, F. Gil-Ortiz, C. Popescu, O. Vallcorba, I. Peral, D. Fullà, J. Benach, and J. Juanhuix, *The Crystallography Stations at the Alba Synchrotron*, Eur. Phys. J. Plus **130**, (2015).

[12] P. J. E. M. Van Der Linden, M. Moretti Sala, C. Henriquet, M. Rossi, K. Ohgushi, F. Fauth, L. Simonelli, C. Marini, E. Fraga, C. Murray, J. Potter, and M. Krisch, *A Compact and Versatile Dynamic Flow Cryostat for Photon Science*, Rev. Sci. Instrum. **87**, (2016).

[13] M. Avdeev and J. R. Hester, *ECHIDNA: A Decade of High-Resolution Neutron Powder Diffraction at OPAL*, J. Appl. Crystallogr. **51**, 1597 (2018).

[14] M. D. Frontzek, R. Whitfield, K. M. Andrews, A. B. Jones, M. Bobrek, K. Vodopivec, B. C. Chakoumakos, and J. A. Fernandez-Baca, *WAND2 - A Versatile Wide Angle Neutron Powder/Single Crystal Diffractometer*, Rev. Sci. Instrum. **89**, (2018).

[15] J. Rodríguez-Carvajal, *Recent Advances in Magnetic Structure Determination by Neutron Powder Diffraction*, Phys. B Phys. Condens. Matter **192**, 55 (1993).

[16] V. D. Juza, G. Dieter, and H. Schäfer, *Uber Die Vanadinjodide $VJ_2$ Und $VJ_3$*, Zeitschrift Fur Anorg. Und Allg. Chemie **322**, (1969).

[17] B. M. De Boisse, M. Reynaud, J. Ma, J. Kikkawa, S. Nishimura, M. Casas-cabanas, C. Delmas, M. Okubo, and A. Yamada, *Coulombic Self-Ordering upon Charging a Large-Capacity Layered*

*Cathode Material for Rechargeable Batteries*, Nat. Commun. **10**, 2185 (2019).

[18]  M. A. McGuire, *Crystal and Magnetic Structures in Layered, Transition Metal Dihalides and Trihalides*, Crystal **7**, 121 (2017).

[19]  S. Jiang, L. Li, Z. Wang, K. F. Mak, and J. Shan, *Controlling Magnetism in 2D CrI$_3$ by Electrostatic Doping*, Nat. Nanotechnol. **13**, 549 (2018).

[20]  T. Li, S. Jiang, N. Sivadas, Z. Wang, Y. Xu, D. Weber, J. E. Goldberger, K. Watanabe, T. Taniguchi, C. J. Fennie, K. Fai Mak, and J. Shan, *Pressure-Controlled Interlayer Magnetism in Atomically Thin CrI$_3$*, Nat. Mater. **18**, 1303 (2019).

[21]  N. Sivadas, S. Okamoto, X. Xu, C. J. Fennie, and D. Xiao, *Stacking-Dependent Magnetism in Bilayer CrI$_3$*, Nano Lett. **18**, 7658 (2018).

[22]  H. T. Stokes, D. M. Hatch, and B. J. Campbell, *ISODISTORT, ISOTROPY Software Suite, Iso.Byu.Edu.*

[23]  B. J. Campbell, H. T. Stokes, D. E. Tanner, and D. M. Hatch, *ISODISPLACE: A Web-Based Tool for Exploring Structural Distortions*, J. Appl. Crystallogr. **39**, 607 (2006).

[24]  P. Jiang, C. Wang, D. Chen, Z. Zhong, Z. Yuan, Z. Y. Lu, and W. Ji, *Stacking Tunable Interlayer Magnetism in Bilayer CrI$_3$*, Phys. Rev. B **99**, 144401 (2018).

Supplementary Information

# Crystallographic and magnetic structures of the VI$_3$ and LiVI$_3$ van der Waals compounds

Thomas Marchandier[1,2,3], Nicolas Dubouis[1,2,3], François Fauth[4], Maxim Avdeev[5,6], Alexis Grimaud[1,2,3], Jean-Marie Tarascon[1,2,3] and Gwenaëlle Rousse[1,2,3]

1. Collège de France, Chaire de Chimie du Solide et de l'Energie, UMR 8260, 11 place Marcelin Berthelot, 75231 Paris Cedex 05, France

2. Réseau sur le Stockage Electrochimique de l'Energie (RS2E), FR CNRS 3459, 75005 Paris, France

3. Sorbonne Université– 4 place Jussieu, F-75005 Paris, France

4. CELLS -ALBA synchrotron, Cerdanyola del Valles, Barcelona E-08290, Spain

5. School of Chemistry, the University of Sydney, Sydney, NSW 2006, Australia

6. Australian Centre for Neutron Scattering, Australian Nuclear Science and Technology Organisation, New Illawarra Rd, Lucas Heights, NSW 2234, Australia

**Part 1:** Structural resolution of the 300 K structure

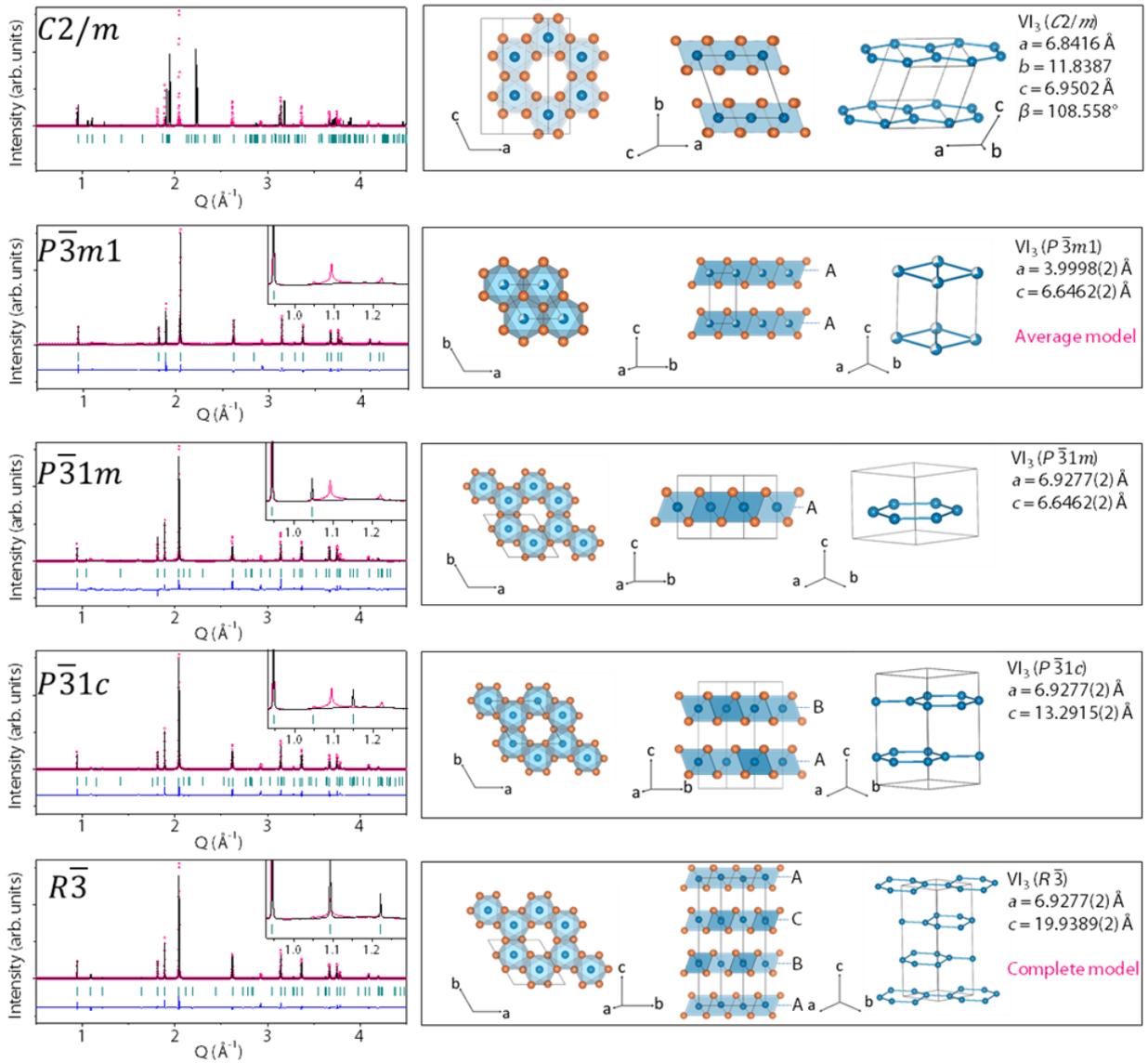

**Figure S1**: Comparison of the Rietveld refinement (left) of the VI$_3$ synchrotron XRD pattern using different structural models with the corresponding structure (right). The inset highlights the superstructure peaks Q region.

**Part 2 :** Structural resolution of the low temperature phases

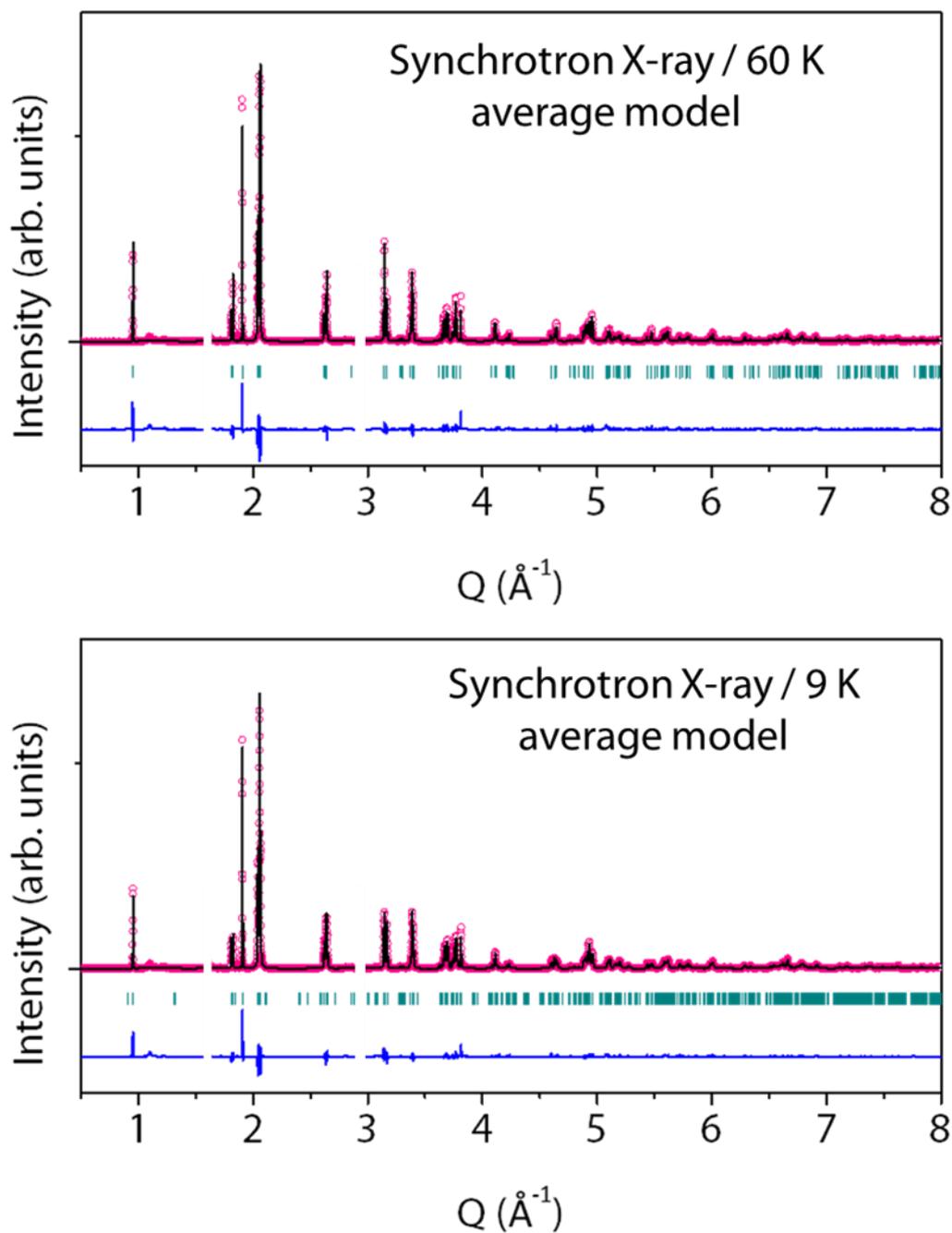

**Figure S2**: Synchrotron XRD Rietveld refinement of $VI_3$ at 60 K (top) and 9 K (bottom). Wavelength for synchrotron X-ray is λ = 0.3869 Å. The pink circles, black continuous line, and bottom blue line represent the observed, calculated, and difference patterns, respectively. Vertical green tick bars stand for the Bragg positions. Some regions have been excluded due to minute amount of impurities.

| VI$_3$ 300 K (complete model) | | $R\bar{3}$ | $R_{Bragg}$ = 15.5 % | $\chi^2$ = 18.5 | | |
|---|---|---|---|---|---|---|
| a = 6.9277(2) Å | | c = 19.9389(2) Å | | | Vol = 828.730(4) Å$^3$ | |
| Atom | Wyckoff Position | x/a | y/b | z/c | $B_{iso}$ (Å$^2$) | Occupancy |
| I | 18f | 0.0115(2) | 0.344(2) | 0.0789(2) | 0.916(11) | 1 |
| V | 6c | 0 | 0 | 0.3266(3) | 0.57(4) | 1 |

**Table S1**: Crystallographic data and atomic positions for VI$_3$ at 300 K using the complete model determined from Rietveld refinement of its synchrotron X-ray pattern.

| VI$_3$ 300 K (average model) | | $P\bar{3}m1$ | $R_{Bragg}$ = 15.0 % | $\chi^2$ = 54.6 | | |
|---|---|---|---|---|---|---|
| a = 3.9998(2) Å | | c = 6.6462(2) Å | | | Vol = 92.081(2) Å$^3$ | |
| Atom | Wyckoff Position | x/a | y/b | z/c | $B_{iso}$ (Å$^2$) | Occupancy |
| I | 2d | 1/3 | 2/3 | 0.2359(2) | 1.031(12) | 1 |
| V | 1a | 0 | 0 | 0 | 1.17(11) | 2/3 |

**Table S2**: Crystallographic data and atomic positions for VI$_3$ at 300 K using the average model determined from Rietveld refinement of its synchrotron X-ray pattern.

2-a) Average model structures

| VI$_3$ 60 K (average model) | | C2/m | $R_{Bragg}$ = 6.94 % | $\chi^2$ = 62.3 | | | |
|---|---|---|---|---|---|---|---|
| a = 6.9395(2) Å | | b = 3.9671(2) Å | c = 6.5960(2) Å | β = 90.4539(4) Å | Vol = 181.578(2) Å$^3$ | | |
| Atom | Wyckoff Position | x/a | y/b | z/c | $B_{iso}$ (Å$^2$) | Occupancy | |
| I | 4i | 0.1669(2) | 1/2 | 0.2402(2) | 0.774(8) | 1 | |
| V | 2a | 0 | 0 | 0 | 0.53(7) | 2/3 | |

**Table S3**: Crystallographic data and atomic positions for VI$_3$ at 60 K using the average model determined from Rietveld refinement of its synchrotron X-ray pattern.

| VI$_3$ 9 K (average model) | P$\bar{1}$ | R$_{Bragg}$ = 8.06 % | $\chi^2$ = 81.3 | | |
|---|---|---|---|---|---|
| a = 6.9353(2) Å | b = 3.9655(2) Å | c = 6.5910(2) Å | | Vol = 181.259(2) Å$^3$ | |
| α = 90.1044(6)° | β = 90.3956(5)° | γ = 90.1495(5)° | | | |
| Atom | Wyckoff Position | x/a | y/b | z/c | B$_{iso}$ (Å$^2$) | Occupancy |
| I1 | 2i | 0.1663(8) | 0.5062(6) | 0.2495(8) | 0.691(9) | 1 |
| I2 | 2i | 0.6668(8) | 0.9945(6) | 0.2373(8) | 0.691(9) | 1 |
| V1 | 1a | 0 | 0 | 0 | 0.48(7) | 2/3 |
| V2 | 1e | 1/2 | 1/2 | 0 | 0.48(7) | 2/3 |

**Table S4**: Crystallographic data and atomic positions for VI$_3$ at 9 K using the average model determined from Rietveld refinement of its synchrotron X-ray pattern.

2-b) Complete model structures

To reintroduce the superstructure to the average model (either at 60 K or 9 K) a first intuitive transformation (transformation 1) can be done to obtain an intermediate cell with metrics close to the $R\bar{3}$ hexagonal cell used at 300 K (cf. figure S3):

$$(a_i \quad b_i \quad c_i) = (a_a \quad b_a \quad c_a) \begin{pmatrix} 1 & -1/2 & 0 \\ 0 & 3/2 & 0 \\ 0 & 0 & 3 \end{pmatrix}$$

Here the i and $a$ subscripts stand for intermediate and average, respectively.

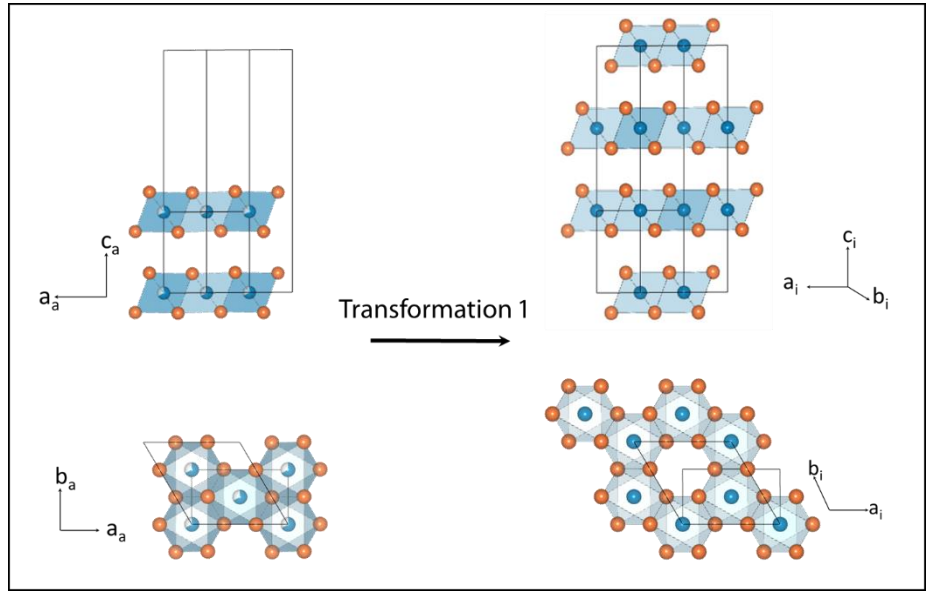

**Figure S3:** Graphical representation of the first transformation. The subscripts "a" and "i" refers to "average" and "intermediatel". For the sake of clarity, both the "average" and "intermediate" unit cell are represented.

In this intermediate cell, it is possible to reconstruct the superstructure following the order observed in the $R\bar{3}$ hexagonal cell at 300 K. Finally, we can show that this intermediate cell is in the $P\bar{1}$ space group equivalent to a smaller cell (refers as supercell) obtained via the following transformation (transformation 2):

$$(a_s \quad b_s \quad c_s) = (a_i \quad b_i \quad c_i) \begin{pmatrix} -1/3 & 2/3 & -1/3 \\ 1/3 & 1/3 & -2/3 \\ 1/3 & 1/3 & 1/3 \end{pmatrix}$$

Here the $s$ subscript stand for supercell.

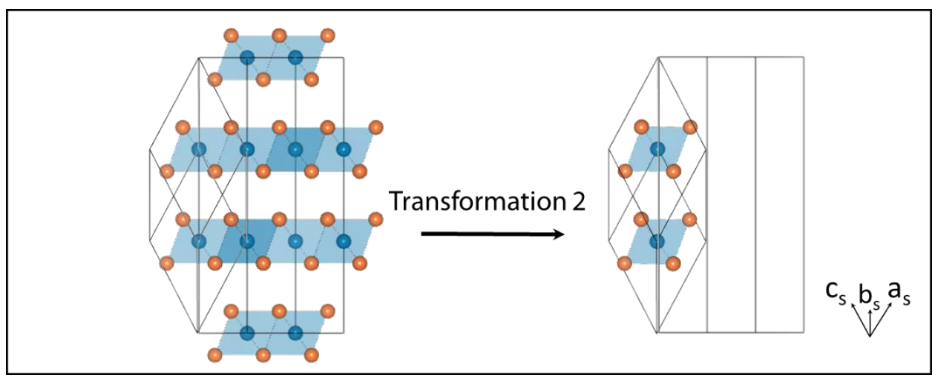

**Figure S4:** Graphical representation of the second transformation. The subscripts "$i$" and "$s$" refers to "intermediate" and "supercell".

| VI$_3$ 60 K (complete model) | | $P\bar{1}$ | $R_{Bragg}$ = 9.42 % | $\chi^2$ = 81.3 | | |
|---|---|---|---|---|---|---|
| $a$ = 7.7359(2) Å | | $b$ = 7.6886(2) Å | $c$ = 7.6968(2) Å | | Vol = 272.356(7) Å$^3$ | |
| $\alpha$ = 53.1933(5)° | | $\beta$ = 53.0168(7)° | $\gamma$ = 53.4732(7)° | | | |
| Atom | Wyckoff Position | x/a | y/b | z/c | $B_{iso}$ (Å$^2$) | Occupancy |
| I1 | 2$i$ | 0.0852(5) | 0.4199(6) | 0.7357(6) | 0.577(8) | 1 |
| I2 | 2$i$ | 0.4157(5) | 0.7471(6) | 0.0771(7) | 0.577(8) | 1 |
| I3 | 2$i$ | 0.7483(5) | 0.0824(6) | 0.4097(7) | 0.577(8) | 1 |
| V1 | 2$i$ | 0.3337(2) | 0.3332(2) | 0.3332(2) | 0.47(6) | 1 |

**Table S5:** Crystallographic data and atomic positions for VI$_3$ at 60 K using the complete model determined from Rietveld refinement of its synchrotron X-ray pattern.

| VI$_3$ 9 K (complete model) | | $P\bar{1}$ | $R_{Bragg}$ = 9.44 % | $\chi^2$ = 86.3 | | |
|---|---|---|---|---|---|---|
| $a$ = 7.7268(2) Å | | $b$ = 7.6808(2) Å | $c$ = 7.6985(2) Å | | Vol = 271.901(3) Å$^3$ | |
| $\alpha$ = 53.1280(2)° | | $\beta$ = 53.0863(3)° | $\gamma$ = 53.5022(3)° | | | |
| Atom | Wyckoff Position | x/a | y/b | z/c | $B_{iso}$ (Å$^2$) | Occupancy |
| I1 | 2$i$ | 0.0822(5) | 0.4148(6) | 0.7438(6) | 0.664(9) | 1 |
| I2 | 2$i$ | 0.4140(5) | 0.7451(5) | 0.0811(6) | 0.664(9) | 1 |
| I3 | 2$i$ | 0.7459(5) | 0.0807(5) | 0.4153(6) | 0.664(9) | 1 |
| V1 | 2$i$ | 0.3335(2) | 0.3335(2) | 0.3334(2) | 0.73(7) | 1 |

**Table S6:** Crystallographic data and atomic positions for VI$_3$ at 9 K using the complete model determined from Rietveld refinement of its synchrotron X-ray pattern.

## Part 3: Magnetic structure of VI$_3$

A simulation of the hypothetical magnetic structures was undertaken to reveal its influence on the NPD pattern. The magnetic framework was built by choosing **k** = (0, 0, 0) as propagation vector (as expected for a ferromagnet) and magnetic moments were placed either perpendicular (cf. figure S5a bottom) or parallel (cf. figure S5a top) to the vanadium honeycomb layers (i.e. along [001] or [100] directions in the $R\bar{3}$ hexagonal cell, respectively). Those two simulated patterns with a magnetic component of 2 μ$_B$, the value expected for a V$^{3+}$ ion (d$^2$), are reported in figure S5b together with a pattern with the nuclear contribution only. These simulations reveal that an orientation of magnetic moments collinearly to the layer plan would induce magnetic contribution to the (003) reflection at Q = 0.9 Å$^{-1}$ (here (hkl) indices refer to the $R\bar{3}$ hexagonal cell), which is not observed experimentally (cf. figure S5b). Contrariwise, an orientation of the moments orthogonally to the layers induces notable magnetic contribution at Q = 1.1 Å$^{-1}$, Q = 1.4 Å$^{-1}$ and Q = 1.6 Å$^{-1}$. If the contribution at Q = 1.6 Å$^{-1}$ can be observed in a much lower amount that expected (cf. figure 7d), the Q = 1.1 Å$^{-1}$, Q = 1.4 Å$^{-1}$ reflection are clearly not visible in the experimental pattern (cf. figure S5b). However, it worth noticing that these two reflections originate from the honeycomb superstructure of the vanadium network. Thus, their absence does not necessarily contradict a long range ordering of the magnetic moment orthogonally to the layer plan but rather originates from local defects and stacking faults.

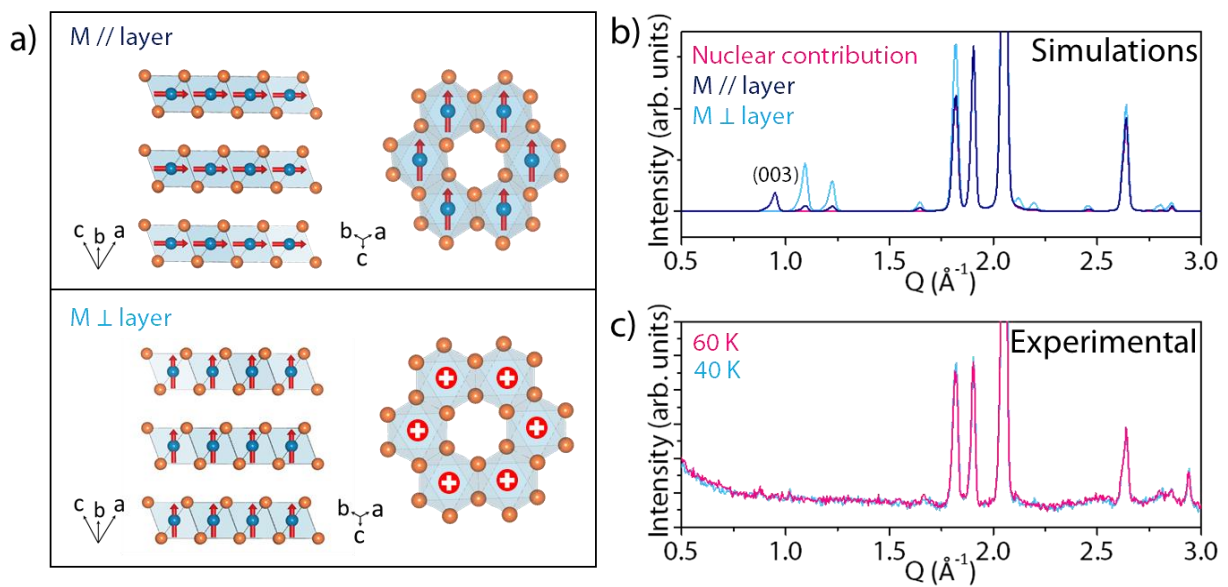

**Figure S5:** a) Magnetic structure of VI$_3$ (vanadium and Iodine atoms are in blue and orange respectively) with spin oriented either parallel (top) or perpendicular (bottom) to the honeycomb layers. The spin orientations are represented with red arrows or by a '+' sign indicating that the spins are pointing orthogonally to the view b) Simulated neutron diffraction patterns with different spin configurations c) comparison of the 60 K and 40 K experimental neutron diffraction pattern in the [0.5 Å$^{-1}$; 3 Å$^{-1}$] Q range.

**Part 4:** LiVI$_3$

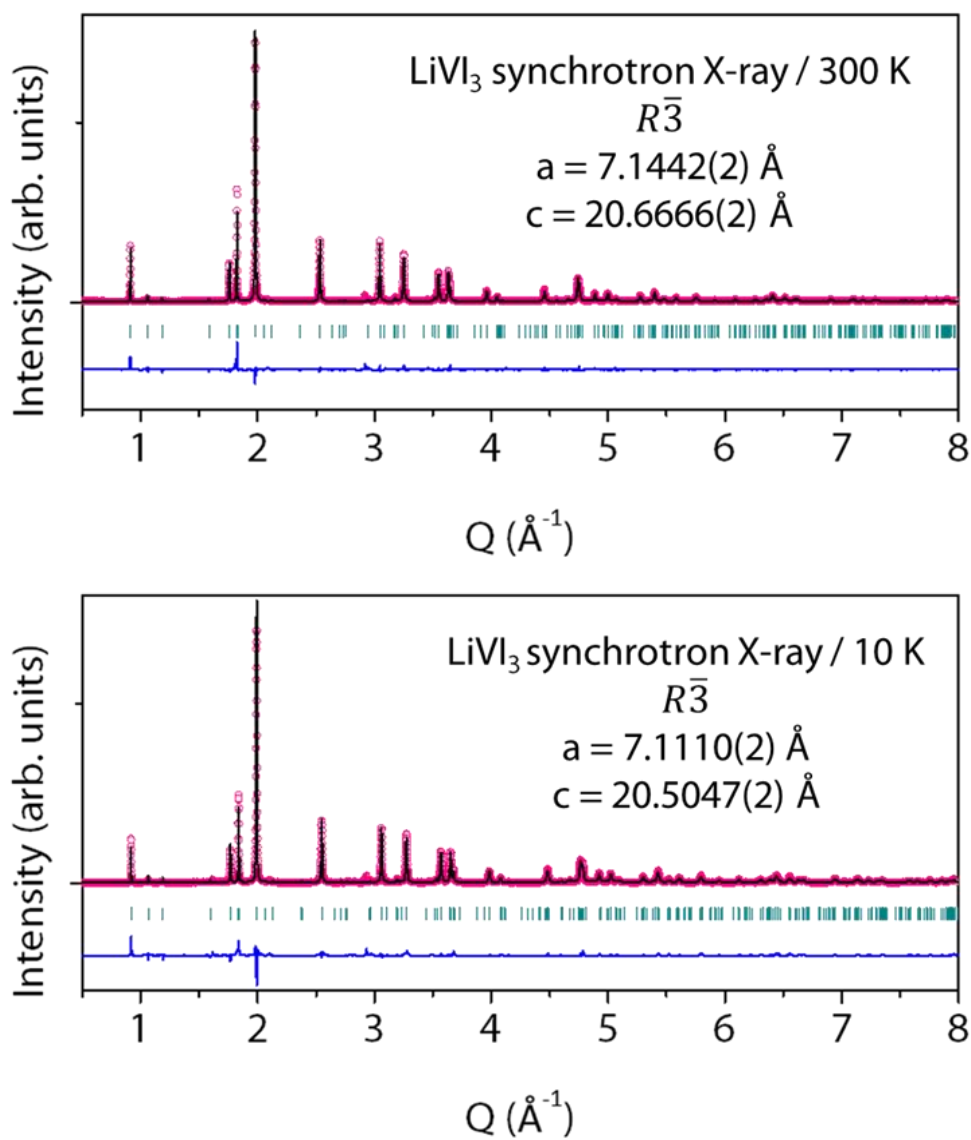

**Figure S6:** Synchrotron XRD Rietveld refinement of LiVI$_3$ at 300 K (top) and 10 K (bottom). Wavelength for synchrotron X-ray is λ = 0.3869 Å. The pink circles, black continuous line, and bottom blue line represent the observed, calculated, and difference patterns, respectively. Vertical green tick bars stand for the Bragg positions.

| LiVI$_3$ 3 K | Shubnikov space group : $R\bar{3}'$ | | R$_{Bragg}$ = 12.0 % | $\chi^2$ = 8.05 | R$_{mag}$ = 16.1 % | | |
|---|---|---|---|---|---|---|---|
| a = 7.0956(2) Å | c = 20.4980(5) Å | | | | Vol = 893.82(4) Å$^3$ | | |
| Atom | Wyckoff Position | x/a | y/b | z/c | B$_{iso}$ (Å$^2$) | Occupancy | M$_z$ |
| I | 18f | 0.0007(14) | 0.328(2) | 0.0792(4) | 0.70(9) | 1 | / |
| V | 6c | 0 | 0 | 0.327(6) | 0.42(4) | 1 | 1.68(6) µ$_B$ |
| Li1 | 3b | 0 | 0 | 1/2 | 2.122(12) | 2/3 | / |
| Li2 | 6c | 0 | 0 | 0.1641(7) | 2.122(12) | 2/3 | / |

**Table S7:** Crystallographic and magnetic structures data and atomic positions for LiVI$_3$ at 3 K determined from Rietveld refinement of its neutron diffraction pattern